\documentclass[structabstract]{aa}
\usepackage{txfonts}
\usepackage{graphicx}
\usepackage{natbib}
\usepackage{longtable}
\usepackage{subeqnarray}
\usepackage{cases}

\usepackage[colorlinks=true,citecolor=blue]{hyperref}

\begin{document}
\title{ALMA view of the $^{12}$C/$^{13}$C isotopic ratio in starburst galaxies}

\author{X. D. Tang\inst{1,2,3}
\and C. Henkel\inst{3,4,1}
\and K. M. Menten\inst{3}
\and Y. Gong\inst{3}
\and S. Mart\'{\i}n\inst{5,6}
\and S. M\"{u}hle\inst{7}
\and S. Aalto\inst{8}
\and S. Muller\inst{8}
\and S. Garc\'{\i}a-Burillo\inst{9}
\and S. A. Levshakov\inst{10}
\and R. Aladro\inst{3}
\and M. Spaans\inst{11}
\and S. Viti\inst{12}
\and H. M. Asiri\inst{4}
\and Y. P. Ao\inst{13}
\and J. S. Zhang\inst{14}
\and X. W. Zheng\inst{15}
\and J. Esimbek\inst{1,2}
\and J. J. Zhou\inst{1,2}
}

\titlerunning{ALMA view of the $^{12}$C/$^{13}$C isotopic ratio in starburst galaxies}
\authorrunning{X. D. Tang et al.}

\institute{
Xinjiang Astronomical Observatory, Chinese Academy of Sciences, 830011 Urumqi, PR China\\
\email{tangxindi@xao.ac.cn}
\and Key Laboratory of Radio Astronomy, Chinese Academy of Sciences, 830011 Urumqi, PR China
\and Max-Planck-Institut f\"{u}r Radioastronomie, Auf dem H\"{u}gel 69, 53121 Bonn, Germany\\
\email{chenkel@mpifr-bonn.mpg.de}
\and Astronomy Department, King Abdulaziz University, PO Box 80203, 21589 Jeddah, Saudi Arabia
\and European Southern Observatory, Alonso de C\'{o}rdova 3107, Vitacura Casilla 763 0355, Santiago, Chile
\and Joint ALMA Observatory, Alonso de C\'{o}rdova 3107, Vitacura Casilla 763 0355, Santiago, Chile
\and Argelander Institut f\"{u}r Astronomie, Universit\"{a}t Bonn, Auf dem H\"{u}gel 71, 53121 Bonn, Germany
\and Department of Earth and Space Sciences, Chalmers University of
Technology, Onsala Observatory, 43992 Onsala, Sweden
\and Observatorio de Madrid, OAN-IGN, Alfonso XII, 3, E-28014-Madrid, Spain
\and Ioffe Physical-Technical Institute, Polytekhnicheskaya Str. 26, 194021 St. Petersburg, Russia
\and Kapteyn Astronomical Institute, University of Groningen, PO Box 800, 9700 AV Groningen, The Netherlands
\and Department of Physics and Astronomy, UCL, Gower St., London, WC1E 6BT, UK
\and Purple Mountain Observatory, Chinese Academy of Sciences, Nanjing 210008, PR China
\and Center for Astrophysics, Guangzhou University, 510006 Guangzhou, PR China
\and School of Astronomy and Space Science, Nanjing University, 210093 Nanjing, PR China
}


\abstract
{We derive molecular-gas-phase $^{12}$C/$^{13}$C isotope ratios for the central few 100\,pc of the three nearby
starburst galaxies NGC\,253, NGC\,1068, and NGC\,4945 making use of the $\lambda$\,$\sim$\,3\,mm
$^{12}$CN and $^{13}$CN\,$N$\,=\,1--0 lines in the ALMA Band 3.
The $^{12}$C/$^{13}$C isotopic ratios derived from the ratios of these lines
range from 30 to 67 with an average of 41.6\,$\pm$\,0.2
in NGC\,253, from 24 to 62 with an average of 38.3\,$\pm$\,0.4 in NGC\,1068,
and from 6 to 44 with an average of 16.9\,$\pm$\,0.3 in NGC\,4945.
The highest $^{12}$C/$^{13}$C isotopic ratios are determined in some of the outskirts of the
nuclear regions of the three starburst galaxies. The lowest ratios are associated with the northeastern and southwestern
molecular peaks of NGC\,253, the northeastern and southwestern edge of the mapped region in
NGC\,1068, and the very center of NGC\,4945. In case of NGC 1068, the measured ratios suggest inflow from
the outer part of NGC\,1068 into the circum-nuclear disk through both the halo and the bar.
Low $^{12}$C/$^{13}$C isotopic ratios in the central regions of these starburst galaxies indicate
the presence of highly processed material.}


\maketitle

\section{Introduction}
\label{sect:Introduction}
Even though interstellar carbon isotope ratios are locally understood
(e.g., \citealt{Wilson1994,Henkel1994b,Wilson1999}), in extragalactic space beyond
the Magellanic Clouds they are almost unexplored. We lack information on
objects outside the Local Group of galaxies tracing environments
that drastically differ from those in the Milky Way and the
Large Magellanic Cloud (LMC). We do not know whether our Galaxy is typical
for its class of objects or whether its isotopic properties are exceptional.
Would they turn out to be peculiar, what would this imply? Moreover,
will we see strong variations
in isotopic ratios when observing nearby galaxies with high angular resolution?

In the past, observational data have been mostly obtained for the Galaxy and
the Magellanic Clouds (e.g., \citealt{Wouterloot1996,Wouterloot2008,Wang2009}).
A surprising result is that the metal-poor outer Galaxy is not merely providing
a “bridge” between the solar neighborhood and the even more metal-poor
LMC. This is explained by the different
age of the bulk of the stellar populations of the outer Galaxy and
the LMC and can be exemplified by the
$^{12}$C/$^{13}$C and $^{18}$O/$^{17}$O ratios, which are both
a measure of “primary” versus “secondary” nuclear processing.
$^{12}$C and $^{18}$O are produced on rapid timescales primarily
via He burning in massive stars. $^{13}$C and $^{17}$O are
predominantly synthesized via CNO processing of $^{12}$C and
$^{16}$O seeds from earlier stellar generations. The latter
occurs on a slower time scale during the red giant phase in low
and intermediate mass stars or novae
(e.g., \citealt{Wilson1994,Henkel1994a,Henkel1994b}).

Molecular spectroscopy is fundamentally important to constrain stellar
nucleosynthesis and the chemical evolution of galaxies. Atomic spectroscopy
of stellar or interstellar gas does not allow us to discriminate between different
isotopic species. However, isotopic abundances are readily obtained by spectroscopy
of molecular isotopologues \citep{Henkel1994b}.
Locally, emphasizing carbon, observational constraints show very
high molecular-gas-phase $^{12}$C/$^{13}$C ratios
($^{12}$C/$^{13}$C given here and elsewhere represent the molecular gas)
from molecular spectroscopy in the outer Galaxy (>100),
high ratios in the local interstellar medium ($\sim$70),
lower ones in the inner Galactic disk and LMC ($\sim$50),
and a smaller value in the Galactic center region ($\sim$20--25)
(e.g., \citealt{Gusten1985,Wilson1994,Henkel1994b,Wouterloot1996,Wilson1999,Wang2009}).
The solar system ratio ($\sim$89; \citealt{Wilson1994,Henkel1994b})
can be interpreted to represent
conditions at a time when the local disk was 4.6$\times$10$^9$ yr
younger than today. Within the framework of “biased infall”
(e.g., \citealt{Chiappini2001}) the Galactic disk is slowly
formed from inside out, which is causing gradients in the
abundances across the disk. The stellar $^{13}$C ejecta,
reaching the interstellar medium with a time delay, are less
dominant in the young stellar disk of the outer Galaxy than
in the inner Galaxy and the older stellar body of the LMC
(for the LMC, see \citealt{Hodge1989}). The solar system ratio,
referring to a younger disk with less $^{13}$C,
is consequently higher than that measured in the present
local interstellar medium.

While the $^{13}$C bearing molecular species can be safely
assumed to be optically thin (with the possible exception of
$^{13}$CO), a basic problem is the optical depth of the $^{12}$C
bearing species (e.g., HCN, HCO$^+$;
\citealt{Nguyen1992,Wild1992,Henkel1993b,Gao2004,Jiang2011,Wang2014,Davis2014,Jimenez2017}).
Beam filling factors, $\tau$($^{12}$CX), are in most
cases unknown when it comes to extragalactic sources.
With this in mind, a useful tracer should possess the following properties:
\begin{itemize}
\item
The tracer must be abundant, showing strong lines, to allow
us to detect also the rare species, albeit not too strong (i.e., opacity $\sim$\,1)
to cause optical thickness problems.
\item
A useful check of the opacity should be provided by transitions exhibiting fine
and hyperfine structure (fs and hfs, respectively). In such cases the splitting should
be wide enough for the different components to be separated for a line emission with
a width of several 100\,km\,s$^{-1}$, commonly encountered in external galaxies.
\item
The components should show, in the optically thin limit, relative intensities as predicted
by local thermodynamical equilibrium (LTE). Then line ratios deviating from these value,
can be used for optical depth estimates.
\item
The tracer must be well understood theoretically and observationally, the former in
terms of its physical and chemical properties related to photodissociation and fractionation, the
latter by a systematic survey of $^{12}$C/$^{13}$C ratios in Galactic star-forming regions.
\item
Finally, there should be no blend with any other potential strong line of another species.
\end{itemize}

Presently, there is only one molecule matching all these conditions,
the cyanide radical (CN). C$_2$H may come close but a systematic Galactic
survey based of this molecule has not yet been conducted.
Previous observations show that CN is widespread in Galactic molecular clouds
(e.g., \citealt{Rodriguez1998,Han2015,Gratier2017,Watanabe2017,Yamagishi2018})
and a variety of other objects, including the circumstellar envelopes of evolved stars
(e.g., \citealt{Bachiller1997,Savage2002,Milam2005,Milam2009,Hily2008,Adande2012}).
CN was detected for the first time in extragalactic sources by \cite{Henkel1988}.
It exhibits strong lines and is therefore also easily detected outside the Galaxy.
(e.g., \citealt{Henkel1988,Henkel1990,Henkel1991,Henkel1993b,Henkel1994a,Henkel1998,Henkel2014,
Aalto2002,Aalto2007,Wang2004,Wang2009,Fuente2005,Perez2007,Perez2009,Garcia2010,
Chung2011,Martin2011,Aladro2013,Aladro2015,Meier2014,Meier2015,Sakamoto2014,
Watanabe2014,Ginard2015,Nakajima2015,Nakajima2018,Saito2015,Konig2016,Qiu2018,Wilson2018}).

CN spectra are complex. Each CN rotational energy level with N\,>\,0
is split into a doublet
by spin-rotation interaction. Because of the spin of the nitrogen
nucleus ($I_1$\,=\,1), each of these components is further split into
a triplet of states. The $^{13}$CN spectrum is further complicated by
the spin of the $^{13}$C nucleus ($I_2$\,=\,1/2). All this results
in a very complex hfs splitting of the rotational lines. Numerically, it has been shown
that carbon ratios resulting from CN measurements should not be affected by isotope
selective photodissociation or chemical fractionation \citep{Langer1984,Roueff2015}.
Observationally, this has been confirmed by \cite{Savage2002} and \cite{Milam2005}.

Previous observations of CN in the Milky Way indicate a $^{12}$C/$^{13}$C isotope ratio gradient
with Galactocentric distance \citep{Savage2002,Milam2005},
which agrees rather well with the gradient derived from measurements of CO and H$_2$CO
(e.g., \citealt{Henkel1980,Henkel1982,Henkel1983,Henkel1985,Henkel1994b,
Langer1990,Langer1993,Wilson1994,Giannetti2014,Yan2019}).
This suggests that the $^{12}$C/$^{13}$C isotope ratios obtained
from CN are an excellent indicator of Galactic chemical evolution \citep{Milam2005}.
This is even more true for extragalactic targets, since CO and
the mm-wave lines of H$_2$CO do not allow for direct determinations of optical
depths of the main species, while the cm-wave lines of H$_2^{13}$CO are extremely weak.

So far, few $^{12}$C/$^{13}$C isotope ratio determinations of extragalactic
targets have been performed. In this paper, we have therefore carried out
observations of three nearby starburst galaxies, NGC\,253, NGC\,1068, and NGC\,4945.
In Sects.\,\ref{sect:Tar-Obs-red} and \ref{sect:Results},
we introduce our targets, observations of CN, data reduction,
and describe the main results. The resulting carbon isotope ratio derived from
CN are then discussed in Sect.\,\ref{sect:discussion}.
Our main conclusions are summarized in Sect.\,\ref{sect:summary}.

\begin{table*}[t]
\caption{Main properties of the galaxies and observational parameters.}
\label{table:Observational_parameters}
\centering
\begin{tabular}
{cccccccc}
\hline\hline 
Source &R.A.(J2000) &DEC.(J2000) &Distance$^a$ &\multicolumn{2}{c}{Beam size} &Position angle &Type$^b$\\
&$^{h}$ {} $^{m}$ {} $^{s}$ &\degr{} \arcmin{} \arcsec &Mpc &"$\times$" &pc$\times$pc &\degr &\\
\hline 
NGC\,253  &00:47:33.14 &--25:17:17.5 &3.9  &3.6$\times$1.7 &$\sim$68$\times$32 &82 &SAB(s)c; Starburst       \\
NGC\,1068 &02:42:40.70 &--00:00:48.0 &14.4 &4.5$\times$2.2 &$\sim$314$\times$154 &75 &(R)SA(rs)b; AGN+Starburst\\
NGC\,4945 &13:05:27.50 &--49:28:06.0 &3.8  &3.6$\times$2.3 &$\sim$66$\times$42 &81 &SB(s)cd; AGN+Starburst   \\
\hline 
\end{tabular}
\tablefoot{$^a$ and $^b$ Distances and type were taken from the NASA/IPAC Extragalactic Database (NED).}
\label{tab:cood}
\end{table*}

\begin{figure*}[t]
\centering
\includegraphics[width=1.00\textwidth]{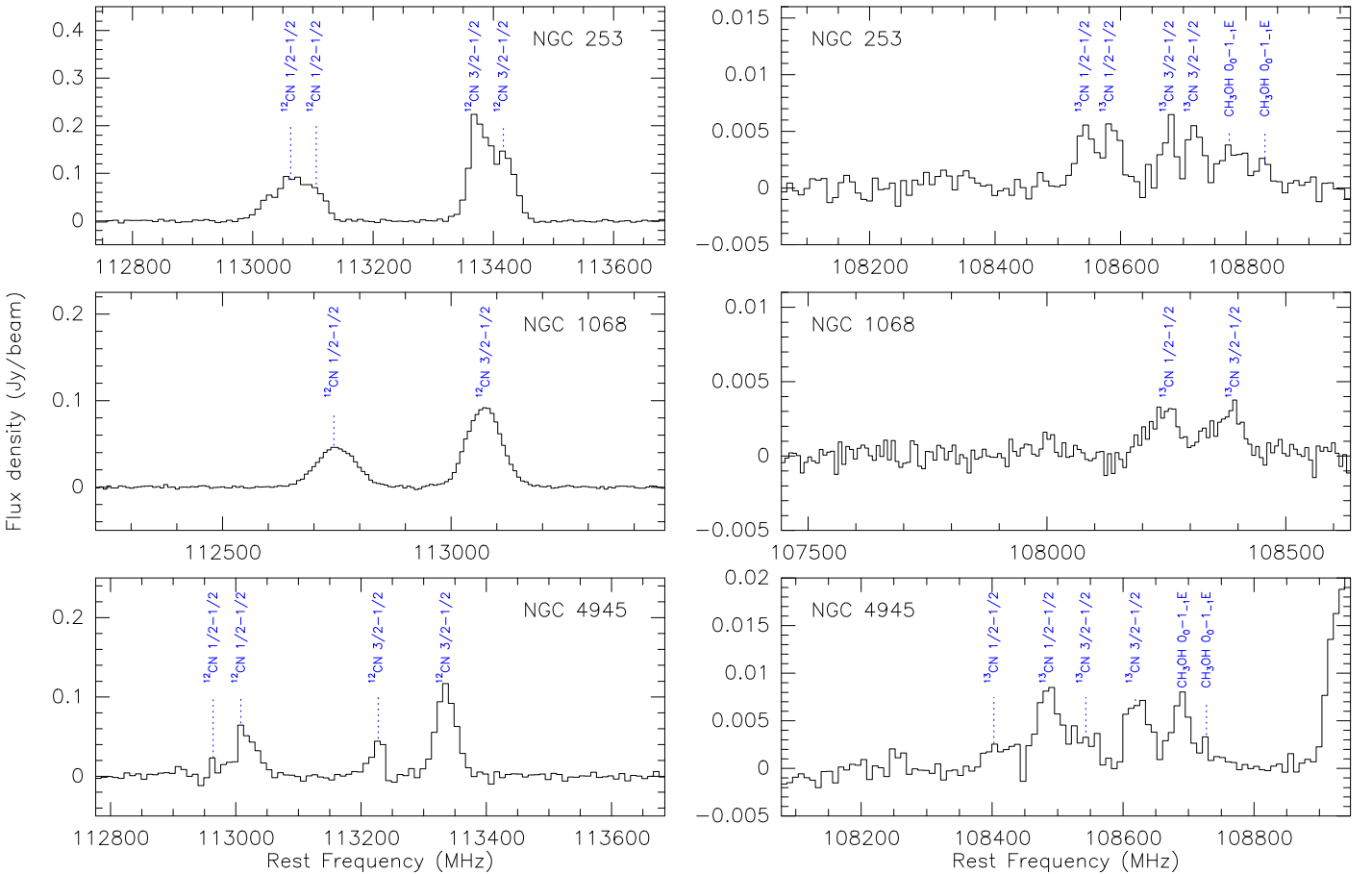}
\caption{$^{12}$CN\,$N$\,=\,1--0 (\emph{left}) and
$^{13}$CN\,$N$\,=\,1--0 (\emph{right}) spectra toward
NGC\,253 (\emph{top}), NGC\,1068 (\emph{middle}),
and NGC\,4945 (\emph{bottom}) obtained with ALMA toward the central position
(see Table\,\ref{tab:cood}).}
\label{fig:spectra}
\end{figure*}

\section{Targets, observations, and data reduction}
\label{sect:Tar-Obs-red}
\subsection{Targets}
\label{sect:Targets}
The three objects we selected, i.e. NGC\,253, NGC\,1068, and NGC\,4945,
are prominent nearby starburst galaxies, exhibiting particularly strong molecular lines
(e.g., \citealt{Martin2006,Chou2007,Garcia2010,Garcia2014,Garcia2016,Garcia2017,Aladro2013,Meier2015,
Henkel2018}). The choice of the galaxies was made to have three of
the strongest extragalactic line emitters and to cover a certain range of starbursts,
i.e. the transition from "moderate starbursts" (NGC\,253 and NGC\,4945) to more luminous
infrared galaxies (LIRGs; NGC\,1068). Single-dish spectral line surveys have been performed for all our selected
starburst galaxies in the 3\,mm band
(e.g., \citealt{Henkel1990,Henkel1994a,Wang2004,Aladro2013,Aladro2015,Nakajima2018}).
The $^{12}$C/$^{13}$C isotope ratio is
$\sim$40 estimated from CN and CS \citep{Henkel1993a,Henkel1993b,Henkel2014},
but $>$81 obtained from C$_2$H \citep{Martin2010}, in the starburst
galaxy NGC\,253. Interferometric measurements of $^{12}$C$^{18}$O/$^{13}$C$^{18}$O
in NGC\,253 indicate a low value of $\sim$21 \citep{Martin2019}.
It is $\sim$50 obtained from CN in NGC\,1068 \citep{Aladro2013}.
A value of 40--50 was reported in NGC\,4945 \citep{Henkel1993a,Henkel1994a,Wang2004}.
Toward M\,82 and IC\,342, \cite{Henkel1998}
found $^{12}$C/$^{13}$C $>$40 and $>$30 from CN.
For Arp\,220 and Mrk\,231 it appears to be 100 from CO and OH \citep{Gonzalez2012,Henkel2014},
and for the Cloverleaf QSO it may be >100 from CO \citep{Henkel2010}.
These determined values indicate a trend matching qualitative expectations
of decreasing $^{12}$C/$^{13}$C values with time and metallicity \citep{Henkel2014}.
However, all classes of sources targeted so far only encompass one or two objects.

Below follows a brief description of the selected targets.
\textbf{NGC\,253}, the Sculptor galaxy, an almost edge-on barred spiral
\citep{Pence1981,Puche1991}, is one of the most
prolific infrared and molecular lighthouses of the entire extragalactic sky.
At a distance of $\sim$3.9\,Mpc (e.g., \citealt{Mouhcine2005,Rekola2005}),
it is a prime example of a galaxy with
a nuclear starburst devoid of an active galactic nucleus
(e.g., \citealt{Ulvestad1997,Henkel2004}). Because of
the exceptional strength of its molecular lines, NGC\,253 was
selected as the target of choice for the first unbiased molecular
line survey of an extragalactic source \citep{Martin2006}. It is
therefore a highly suitable target for this study.\\
\textbf{NGC\,1068}, a local LIRG and a prototypical
Seyfert 2 galaxy with a starburst at a distance of $\sim$14.4\,Mpc \citep{Bland1997},
is one of the best extragalactic targets for studying the physical and chemical
properties of the interstellar medium (ISM) in the vicinity of an active galactic nucleus (AGN).
Numerous molecular line observations have targetted NGC\,1068's circumnuclear
molecular ring and the effect of nuclear activity on its ISM
(e.g., \citealt{Schinnerer2000,Usero2004,Garcia2010,Garcia2014,Garcia2016,Garcia2017,
Krips2011,Aladro2013,Viti2014,Wang2014,Qiu2018}). H$_2$O maser emission has been
observed forming an edge-on disk in the circumnuclear environment of NGC\,1068
(e.g., \citealt{Greenhill1996,Gallimore1996,Gallimore1997,Gallimore2001}).\\
\textbf{NGC\,4945}, at a distance of $\sim$3.8\,Mpc
(e.g., \citealt{Karachentsev2007,Mould2008}), has an active Seyfert 2 nucleus and is,
like NGC\,253, an almost edge-on spiral galaxy.
Its central region is known to show
a rich molecular spectrum hosting not only a nuclear starburst
but also an AGN (e.g., \citealt{Marconi2000,Yaqoob2012}).
Past molecular single-dish and interferometric observational studies exist for
CO, CS, CN, HCN, HNC, HCO$^+$, CH$_3$OH, H$_2$CO
(e.g., \citealt{Henkel1990,Henkel1994a,Henkel2018,
Dahlem1993,Mauersberger1996,Curran2001,Wang2004,Chou2007,Hitschfeld2008,Green2016,McCarthy2018}).
H$_2$O megamaser emission has been observed in the nucleus of NGC\,4945 \citep{Greenhill1997}.

\subsection{Observations}
\label{sect:Observations}
Our observations were carried out from 2014 December to 2015 January with
the Atacama Large Millimeter/submillimeter Array (ALMA) in Band 3 (Project:\,2013.1.01151.S).
During the observations, 36--40 12-m antennas were employed in a compact configuration
with baselines ranging from 15 to 349\,m.  For each source, the observations took about 30 minutes
and 1 hour for $^{12}$CN and $^{13}$CN, respectively.
The CN\,$N$\,=\,1--0 transition consists of 9 hyperfine
components blended into two groups, with the stronger group representing
the $J$\,=\,3/2--1/2 transitions and the weaker group the
$J$\,=\,1/2--1/2 transitions. The $^{12}$CN\,$N$\,=\,1--0 ($J$\,=\,1/2--1/2 and 3/2--1/2)
and $^{13}$CN\,$N$\,=\,1--0 ($J$\,=\,1/2--1/2 and 3/2--1/2)
transitions have intensity weighted rest frequencies of 113.191, 113.491, 108.658,
and 108.780\,GHz,
respectively, when LTE line ratios and optically thin emission are adopted.
On each, the $^{12}$CN and the $^{13}$CN line, a spectral window was centered,
with a bandwidth of 1875\,MHz and a frequency resolution of 7812.5\,kHz, corresponding to
a channel width of $\sim$21\,km\,s$^{-1}$.
Basic observational parameters as well as phase center coordinates are listed in
Table\,\ref{table:Observational_parameters}. The $^{12}$CN data of NGC\,253
were not part of our project. Instead, \cite{Meier2015} observed NGC\,253 in $^{12}$CN\,$N$\,=\,1--0 with
ALMA Band 3 (Project:\,2011.0.00172.S). Quality and, in particular, beam sizes
of these data are well matching our ALMA observations.
$^{12}$CN\,$N$\,=\,1--0 data of NGC\,253 are therefore taken from \cite{Meier2015}
in this work. The observed spectra toward the central positions of NGC\,253, NGC\,1068, and
NGC\,4945 are shown in Fig.\,\ref{fig:spectra}.

\subsection{Data reduction}
\label{sect:reduction}
The data were calibrated and imaged using the
CASA\footnote{\tiny https://casa.nrao.edu} 4.2 version pipeline \citep{McMullin2007}.
Our $^{13}$CN data reduction for NGC\,253 (see Table\,\ref{table:Observational_parameters})
follows that performed for the $^{12}$CN data
set (beam size $\sim$3.8$''$$\times$2.7$''$ and position angle --31$\degr$)
by \cite{Meier2015}. The spectral line images were analysed using
GILDAS\footnote{\tiny http://www.iram.fr/IRAMFR/GILDAS}.
Typical rms noise levels in the $^{12}$CN and $^{13}$CN velocity-integrated intensity images
are 1--2 and 0.3--0.7\,mJy\,beam$^{-1}$ for NGC\,253, 1--2 and
0.4--0.7\,mJy\,beam$^{-1}$ for NGC\,1068, and 2--4 and
0.4--0.9\,mJy\,beam$^{-1}$ for NGC\,4945, respectively.

\subsection{Spatial filtering}
\label{sect:Spatial-filtering}
As mentioned in Sect.\,\ref{sect:Targets}, the CN\,$N$\,=\,1--0 transitions has previously
been observed using the 15-m SEST (beam size $\sim$44$''$), 30-m IRAM (beam
size $\sim$22$''$), and 45-m NRO (beam size $\sim$15$''$) telescopes
toward NGC\,253, NGC\,1068, and NGC\,4945
\citep{Henkel1990,Henkel1994a,Henkel2014,Wang2004,Aladro2013,Aladro2015,Nakajima2018}.
From these observations $^{12}$C/$^{13}$C isotope ratios of 40$\pm$10, $\sim$50,
and 40--50, respectively, were determined.
One should note that the $^{13}$CN\,$N$\,=\,1--0 features observed with
the 30-m IRAM telescope by \cite{Aladro2013} are weak and
show low signal-to-noise ratios.
Both $^{12}$CN and $^{13}$CN\,$N$\,=\,1--0 lines of NGC\,1068 were detected with slightly
higher signal-to-noise ratios with the 45-m NRO telescope \citep{Nakajima2018}.
A $^{12}$C/$^{13}$C isotope ratio of $\sim$40 in NGC\,1068 in calculated from Nakajima et al.'s
integrated intensity ratio of $I$($^{12}$CN)/$I$($^{13}$CN) following our method mentioned
in Sect.\,\ref{sect:ISO}. For NGC\,4945, the $^{13}$CN\,$N$\,=\,1--0 transition was not
detected by \cite{Wang2004}.

Due to spatial filtering, the missing flux, that is flux of large scale structures
not sampled by the interferometer, may affect $^{12}$CN/$^{13}$CN line ratios.
To evaluate the missing flux we reconstruct our ALMA data with beams of $\sim$22$''$
(IRAM 30-m; \citealt{Henkel2014,Aladro2015}), $\sim$15$''$ (NRO 45-m; \citealt{Nakajima2018}),
and $\sim$44$''$ (SEST 15-m; \citealt{Wang2004}) for NGC\,253, NGC\,1068, and NGC\,4945,
respectively. We find that $\sim$89\% and $\sim$86\% of the $^{12}$CN and $^{13}$CN
integrated flux observed by single-dish telescope is recovered for NGC\,253 by our ALMA data, respectively.
Only $\sim$30\% and $\sim$53\% of the $^{12}$CN and $^{13}$CN single-dish integrated
flux of NGC\,1068 is recovered, respectively. However, as mentioned above, the
$^{13}$CN\,$N$\,=\,1--0 features observed with the 45-m NRO telescope
\citep{Nakajima2018} are weak and show large uncertainties. For NGC\,4945,
$\sim$63\% of the $^{12}$CN single-dish integrated flux is recovered.
We do not evaluate the missing flux of $^{13}$CN in NGC\,4945
because (as already mentioned) the $^{13}$CN\,$N$\,=\,1--0 transition was not detected
with the 15-m SEST \citep{Wang2004}. In their Table 2, \cite{Wang2004} provide
an upper limit (3 sigma) to the $^{13}$CN\,$N$\,=\,1--0 line intensity,
so at least $\sim$40\% is recovered by our ALMA data. Based on similar missing flux
of our $^{12}$CN and $^{13}$CN data in NGC\,253 and NGC\,1068, we note that
$^{13}$CN\,$N$\,=\,1--0 is likely showing a less extended morphology
(see Sect.\,\ref{sect:CN-distribution}) only because it is more rapidly reaching intensities
below the detection threshold outside the line peaks. So our $^{13}$CN data may cover a
single-dish integrated flux fraction which is similar to that of $^{12}$CN in NGC\,4945.
With similar missing flux levels related to our $^{12}$CN and $^{13}$CN data in
NGC\,253 and NGC\,1068, which is based on a comparison of our interferometric measurements
with previously published single-dish observations, missing flux appears to affect
$^{12}$CN/$^{13}$CN line ratios in NGC\,253, NGC\,1068, and NGC\,4945 only weakly.

There may be line blending with the CH$_3$OH (0$_0$--1$_{-1}$\,E) transition
at 108.894\,GHz, 114\,MHz ($\sim$300\,km\,s$^{-1}$) off the group of
$^{13}$CN\,($N$\,=\,1--0; $J$\,=\,3/2--1/2) lines centered at 108.780\,GHz
(see Sect.\,\ref{sect:Tar-Obs-red} and Fig.\,\ref{fig:spectra}).
However, CH$_3$OH can be well identified in most locations of NGC\,253 and NGC\,4945,
which indicates that the $I$($^{12}$CN)/$I$($^{13}$CN) ratios
are at most only weakly affected by CH$_3$OH in these two galaxies.
For NGC\,1068, the typical linewidths of $^{13}$CN are broader ($\sim$170\,km\,s$^{-1}$)
and CH$_3$OH can not be identified in most locations.
This could indicate that the $I$($^{12}$CN)/$I$($^{13}$CN) ratio
may be underestimated from our ALMA data in NGC\,1068.
The slightly less extended $^{13}$CN distributions also addressed in Sect.\,\ref{sect:CN-distribution}
when compared to $^{12}$CN may merely indicate that the minimum detectable
molecular H$_2$ column density
is higher than in case of $^{12}$CN with its higher fractional abundance
(see also Sect.\,\ref{sect:opacity}).

\begin{figure*}[t]
\centering
\includegraphics[width=1.00\textwidth]{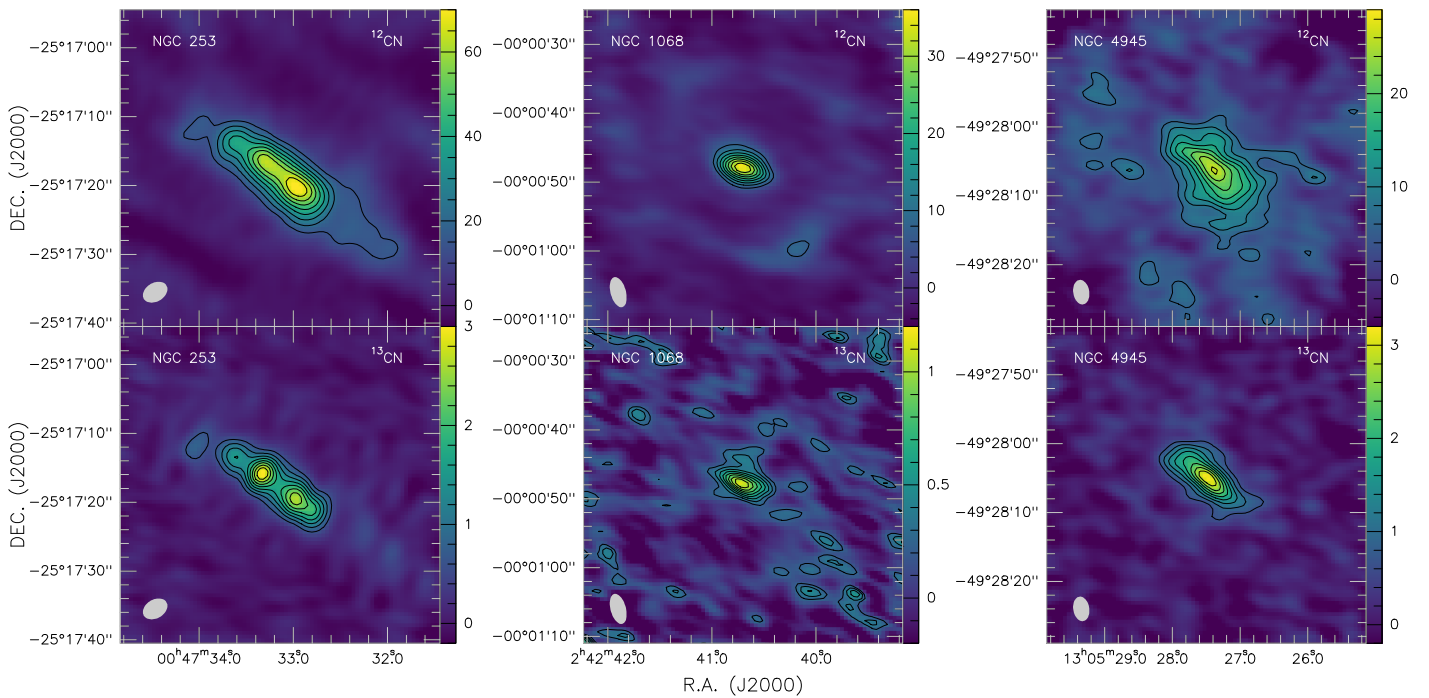}
\caption{Integrated intensity maps (color bars in units of
Jy\,beam$^{-1}$\,km\,s$^{-1}$) of $^{12}$CN ($I$($^{12}$\rm CN\,$J$\,=\,1/2--1/2)+$I(^{12}$\rm CN\,$J$\,=\,3/2--1/2))
and $^{13}$CN ($I$($^{13}$\rm CN\,$J$\,=\,1/2--1/2)+$I(^{13}$\rm CN\,$J$\,=\,3/2--1/2))
of NGC\,253 ($^{12}$CN and $^{13}$CN integrated frequency range: 112.960--113.491 and
108.526--108.744\,GHz; \emph{left}), NGC\,1068 ($^{12}$CN and $^{13}$CN integrated
frequency range: 112.641--113.169 and 108.170--108.459\,GHz; \emph{middle}),
and NGC\,4945 ($^{12}$CN and $^{13}$CN integrated frequency range: 112.809--113.377
and 108.363--108.653\,GHz; \emph{right}). The contour levels are from 20\%
to 100\% with steps of 10\% for $^{12}$CN and $^{13}$CN
of the peak intensity. The $^{12}$CN peak intensities are 71.9, 35.8, and 26.4
Jy\,beam$^{-1}$\,km\,s$^{-1}$, and the $^{13}$CN peak intensities are 3.0, 1.1,
and 3.2 Jy\,beam$^{-1}$\,km\,s$^{-1}$ in NGC\,253, NGC\,1068, and NGC\,4945, respectively.
For the $^{12}$CN map of NGC\,253 also see \cite{Meier2015}.
The pixel size of each image is 0.3$''$$\times$0.3$''$.
The synthesized beam of each image is shown in the lower left corner.}
\label{fig:CN-maps}
\end{figure*}

\section{Results}
\label{sect:Results}
\subsection{Distribution of $^{12}$CN and $^{13}$CN}
\label{sect:CN-distribution}
The integrated intensity distributions of $^{12}$CN and $^{13}$CN
in NGC\,253, NGC\,1068, and NGC\,4945 are shown in Fig.\,\ref{fig:CN-maps}.
In all three galaxies $^{12}$CN shows extended distributions,
in agreement with previous observations of $^{12}$CN in galaxies
(e.g., \citealt{Henkel1988,Garcia2010,Meier2014,Sakamoto2014,Ginard2015,Nakajima2015,
Saito2015,Wilson2018}).
This is also consistent with previous observational results for
other gas tracers as e.g., $^{13}$CO, C$^{18}$O, CS, HCN, or HCO$^+$ toward our sources
(e.g., \citealt{Krips2011,Sakamoto2011,Garcia2014,Viti2014,Meier2015,Henkel2018,Tan2018,Martin2019}).
Weak $^{12}$CN emission is detected in the spiral arms of NGC\,1068
(see Fig.\,\ref{fig:CN-maps}). $^{13}$CN is only detected in the central
regions of NGC\,253, NGC\,1068, and NGC\,4945, and shows slightly
less extended distributions than $^{12}$CN (see also the end of Sect.\,\ref{sect:Spatial-filtering}).

\subsection{$^{12}$CN/$^{13}$CN line ratios}
\label{sect:line-ratio}
\begin{figure*}[t]
\vspace*{0.2mm}
\begin{center}
\includegraphics[width=1.00\textwidth]{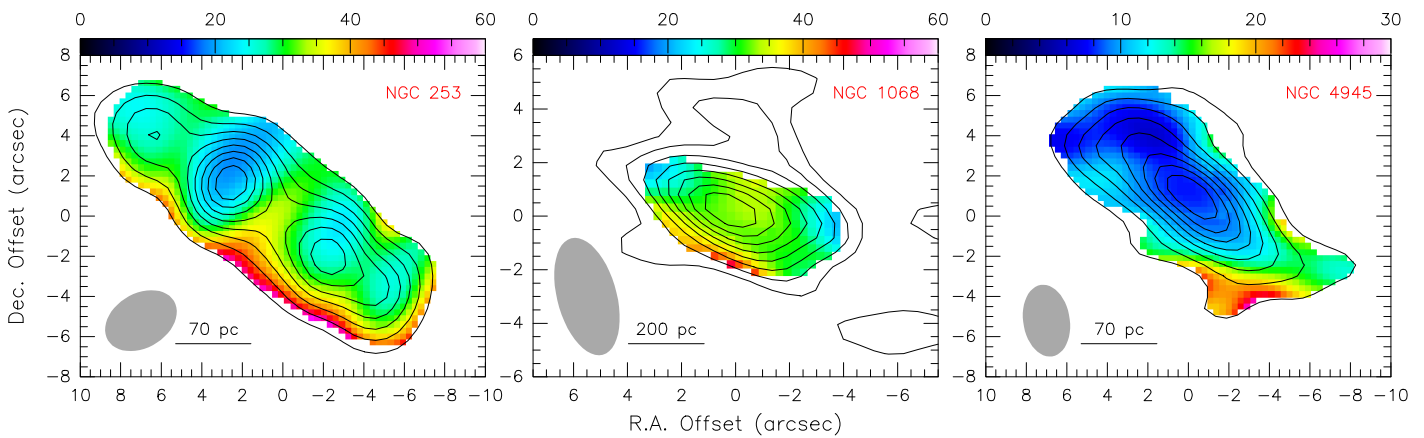}
\end{center}
\caption{Velocity-integrated intensity ratio maps of
$I$($^{12}$CN)/$I$($^{13}$CN) (see Sects.\,\ref{sect:line-ratio}
and \ref{sect:ISO}) in NGC\,253 (\emph{left}), NGC\,1068 (\emph{middle}),
and NGC\,4945 (\emph{right}). Black contours delineate levels of $^{13}$CN integrated
intensity (same as in Fig.\,\ref{fig:CN-maps}).}
\label{fig:CN-ratios}
\end{figure*}

$^{12}$CN/$^{13}$CN line ratio maps of NGC\,253, NGC\,1068, and NGC\,4945 are shown in
Fig.\,\ref{fig:CN-ratios}.
The line ratios are calculated using velocity-integrated intensities where
the $^{13}$CN lines are detected with signal-to-noise ratios of S/N\,$\gtrsim$\,5$\sigma$.
$I$($^{12}$CN)/$I$($^{13}$CN) ratios range from
19 to 53 with an average of 30.0\,$\pm$\,0.2
(errors given here and elsewhere are standard deviations of the mean) in NGC\,253,
from 20 to 47 with an average of 31.4\,$\pm$\,0.3 in NGC\,1068, and from 6 to 25
with an average of 11.5\,$\pm$\,0.1 in NGC\,4945. High ratios ($>$30)
are obtained in the outskirts of the region analyzed by us in NGC\,253.
It shows that low ratios ($<$30) associate with $^{13}$CN peak emission in NGC\,253.
Gradients are seen from the center to the northeastern and to the southwestern
region of NGC\,1068. Two locations in the northeast and southwest
(see Table\,\ref{table:Opacity-peak} and Sect.\,\ref{sect:Variation}) have a low ratio (<25).
For NGC\,4945, low ratios ($<$15) associate with $^{13}$CN peak emission and the northeastern
region and high ratios ($>$15) are located in the outskirts and the southwestern region.
One should note that the $^{13}$CN emission is weak and
shows low signal-to-noise ratios in the edge regions of our three targets.
This may lead to large uncertainties of $I$($^{12}$CN)/$I$($^{13}$CN) ratios in these locations.

\begin{figure*}[t]
\vspace*{0.2mm}
\begin{center}
\includegraphics[width=1.00\textwidth]{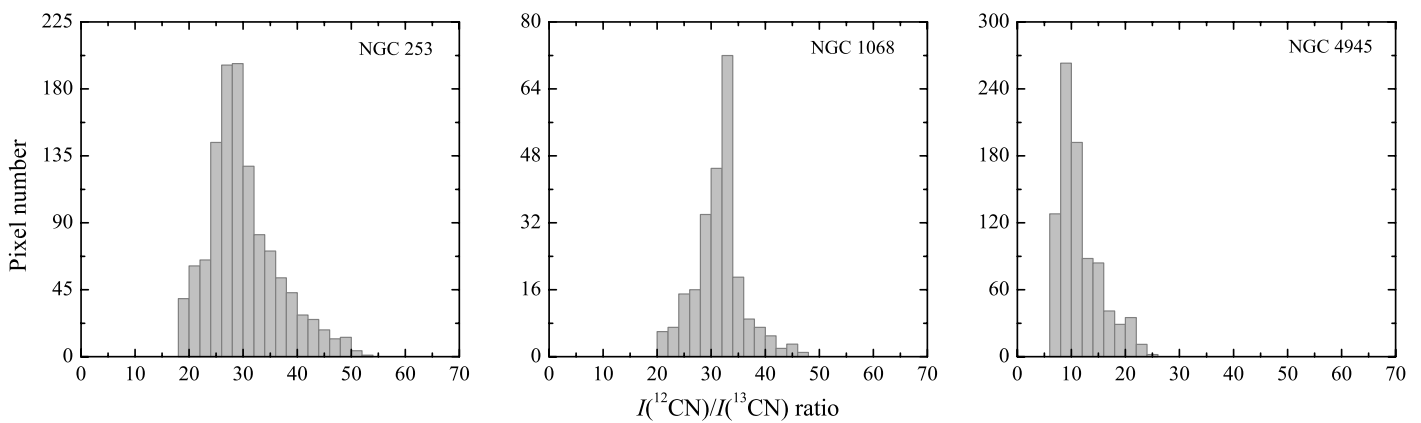}
\end{center}
\caption{Histograms showing the velocity-integrated intensity
ratios of $I$($^{12}$CN)/$I$($^{13}$CN) for NGC\,253 (\emph{left}),
NGC\,1068 (\emph{middle}), and NGC\,4945 (\emph{right}).}
\label{fig:CN-ratios-hisgram}
\end{figure*}

\subsection{Opacities of $^{12}$CN and $^{13}$CN}
\label{sect:opacity}
Previous observations toward external galaxies suggest that optical depths of
the $^{12}$CN\,$N$\,=\,1--0 transition are modest
(e.g., \citealt{Henkel1998,Henkel2014,Wang2009,Meier2015,Konig2016,Nakajima2018}).
For the $N$\,=\,1--0 line of the rare $^{13}$CN isotopologue this holds of course as well.
Opacities can be determined by an analysis of line intensity ratios
because (as already mentioned in Sect.\,\ref{sect:Introduction})
the $N$\,=\,1--0 transition is split into
several spectral features. If $^{12}$CN\,$N$\,=\,1--0 is optically thin
and LTE prevails, line intensity ratios should be 2:1
for $^{12}$CN\,(3/2--1/2)/(1/2--1/2) and 1:1.225 for $^{13}$CN\,(3/2--1/2)/(1/2--1/2)
(see Fig.\,\ref{fig:spectra}). Previous single-dish observations of
$^{12}$CN\,$N$\,=\,1--0 in galaxies indicate that the $^{12}$CN\,(3/2--1/2)/(1/2--1/2)
line intensity ratio is indeed $\sim$2 (e.g.,
\citealt{Henkel1998,Henkel2014,Aladro2013,Aladro2015,Watanabe2014,Nakajima2018}).

We averaged all pixels in NGC\,253, NGC\,1068, and NGC\,4945 for which the $^{13}$CN
line is detected with signal-to-noise ratios of S/N\,$\gtrsim$\,5$\sigma$.
The line integrated intensity ratios of the two $^{12}$CN\,$N$\,=\,1--0 features are
1.58\,$\pm$\,0.11, 1.70\,$\pm$\,0.02, and 1.68\,$\pm$\,0.25
in NGC\,253, NGC\,1068, and NGC\,4945, respectively, which is
slightly lower than that for LTE and optically thin emission. Following the method
applied by \cite{Wang2004}, the optical depth of the $^{12}$CN\,$N$\,=\,1--0 lines
can be obtained from the integrated intensity ratio of $^{12}$CN\,(3/2--1/2)/(1/2--1/2) as
\begin{equation}
\frac{I(^{12}{\rm CN}\,J\!=\!3/2\!-\!1/2)}{I(^{12}{\rm CN}\,J\!=\!1/2\!-\!1/2)}
=\frac{1\!-\!{\rm e}^{-\tau_1}}{1\!-\!{\rm e}^{-\tau_2}},
\label{Equ:opa}
\end{equation}
where $\tau_1$ and $\tau_2$ are the optical depths of
$^{12}$CN\,$J$\,=\,3/2--1/2 and 1/2--1/2, respectively, and $\tau_1$\,=\,2$\tau_2$
(see \citealt{Skatrud1983} for relative LTE intensities under optically
thin conditions). From the above average integrated intensity ratios of
the two $^{12}$CN\,$N$\,=\,1--0 features the derived optical depths of
$^{12}$CN\,$J$\,=\,3/2--1/2 are 1.1, 0.7, and 0.8 in NGC\,253, NGC\,1068,
and NGC\,4945, respectively. Opacities of $^{12}$CN at $^{13}$CN peaks and
outskirts are also calculated in Table\,\ref{table:Opacity-peak}. These suggest that the opacity of
$^{12}$CN\,$N$\,=\,1--0 slightly affects the line intensity ratios of $^{12}$CN/$^{13}$CN
in our three selected starburst galaxies. For $^{13}$CN\,$N$\,=\,1--0, the corresponding
average line integrated intensity ratios are 1.04\,$\pm$\,0.08, 1.03\,$\pm$\,0.14,
and 1.25\,$\pm$\,0.01, which is consistent with LTE and optically thin emission
considering the uncertainties of our interferometric measurements.
In this work we assume that the $^{13}$CN\,$N$\,=\,1--0 line is optically thin
in all of our three studied objects.

\begin{table*}[t]
\caption{$^{12}$CN/$^{13}$CN line ratios, $^{12}$CN opacities, and the $^{12}$C/$^{13}$C isotope ratios toward the $^{13}$CN peaks and outskirts.}
\label{table:Opacity-peak}
\centering
\begin{tabular}
{cccccccccccc}
\hline\hline 
Source &Offset &$I$($^{12}$CN)/$I$($^{13}$CN) &$I(^{12}$CN\,3/2--1/2)/$I(^{12}$CN\,1/2--1/2) &$\tau_1(^{12}$CN 3/2--1/2) &$^{12}$C/$^{13}$C & Note\\
\hline 
NGC\,253 &(6.3\arcsec, 3.9\arcsec)    &26.4$\pm$1.2 &1.67$\pm$0.03 &0.79$\pm$0.08 &33.1$\pm$3.0  &Peak     \\
         &(2.7\arcsec, 1.5\arcsec)    &20.0$\pm$0.5 &1.49$\pm$0.01 &1.42$\pm$0.06 &31.1$\pm$1.6  &Peak     \\
         &(--2.4\arcsec, --2.1\arcsec)&25.7$\pm$1.1 &1.55$\pm$0.01 &1.21$\pm$0.05 &37.2$\pm$2.6  &Peak     \\
         &(--4.5\arcsec, --3.6\arcsec)&27.3$\pm$2.2 &1.53$\pm$0.02 &1.28$\pm$0.06 &40.6$\pm$4.6  &Peak     \\
         &(3.9\arcsec, 5.7\arcsec)    &30.3$\pm$4.0 &1.50$\pm$0.05 &1.40$\pm$0.21 &46.7$\pm$10.6 &Outskirts\\
         &(--1.5\arcsec, --5.1\arcsec)&52.7$\pm$4.4 &1.66$\pm$0.02 &0.82$\pm$0.07 &66.9$\pm$8.8  &Outskirts\\
NGC\,1068&(0.3\arcsec, 0.3\arcsec)    &32.9$\pm$2.7 &1.77$\pm$0.02 &0.52$\pm$0.05 &37.3$\pm$4.3  &Peak     \\
         &(0.3\arcsec, --1.8\arcsec)  &47.2$\pm$8.0 &1.63$\pm$0.03 &0.93$\pm$0.11 &62.2$\pm$16.4 &Outskirts\\
         &(--0.6\arcsec, 1.2\arcsec)  &35.3$\pm$3.3 &1.60$\pm$0.09 &1.03$\pm$0.33 &48.2$\pm$13.2 &Outskirts\\
         &(3.3\arcsec, 1.8\arcsec)    &20.0$\pm$2.3 &1.72$\pm$0.05 &0.66$\pm$0.14 &24.0$\pm$4.9  &Outskirts\\
         &(--3.6\arcsec, --0.3\arcsec)&21.0$\pm$3.0 &1.67$\pm$0.06 &0.80$\pm$0.18 &26.5$\pm$7.1  &Outskirts\\
NGC\,4945&(0.0\arcsec, 0.9\arcsec)    &7.9$\pm$0.3  &1.78$\pm$0.10 &0.50$\pm$0.27 &9.0$\pm$1.7   &Peak     \\
         &(--3.0\arcsec, --4.5\arcsec)&26.6$\pm$3.6 &1.39$\pm$0.11 &1.86$\pm$0.65 &43.6$\pm$20.8 &Outskirts\\
         &(--4.8\arcsec, --0.3\arcsec)&18.9$\pm$2.2 &1.39$\pm$0.11 &1.88$\pm$0.69 &34.0$\pm$15.1 &Outskirts\\
\hline 
\end{tabular}
\tablefoot{For the reference positions to the offsets given in column 2,
see Table\,\ref{table:Observational_parameters}. Calibration uncertainties are not
considered here but are discussed in Sect.\,\ref{sect:ISO}.}
\label{tab:averaged-ratio}
\end{table*}

\subsection{The $^{12}$C/$^{13}$C isotopic ratios from CN}
\label{sect:ISO}
As mentioned in Sect.\,\ref{sect:opacity}, the $^{12}$CN\,$N$\,=\,1--0
line is slightly influenced by saturation
effects in the starburst galaxies NGC\,253, NGC\,1068, and NGC\,4945.
Therefore, in these objects this line's intensities are not exactly proportional
to CN column densities.
Nevertheless, the integrated intensity ratios of $^{12}$CN and $^{13}$CN can
be used to determine the $^{12}$C/$^{13}$C isotope ratios.
Assuming the $^{13}$CN\,$N$\,=\,1--0 lines are optically
thin and assuming that LTE holds in NGC\,253, NGC\,1068, and NGC\,4945 and
following the method applied by \cite{Henkel2014}, the $^{12}$C/$^{13}$C
isotope ratio can be obtained from the integrated intensity ratio
of $^{12}$CN/$^{13}$CN\,$N$\,=\,1--0 as
\begin{equation}
\frac{^{12}\rm C}{^{13}\rm C}\!\approx\!\frac{f_1\!\times\!I(^{12}{\rm CN}\,
J\!=\!3/2\!-\!1/2)\!+\!f_2\!\times\!I(^{12}{\rm CN}\,J\!=\!1/2\!-\!1/2)}
{1.082\!\times\!I(^{13}\rm CN)},
\label{Equ:1}
\end{equation}
where $f_1$\,=\,$\frac{\tau_1}{1-{\rm e}^{-\tau_1}}$ and
$f_2$\,=\,$\frac{\tau_2}{1-{\rm e}^{-\tau_2}}$, and $\tau_1$ and $\tau_2$
are the optical depths of $^{12}$CN\,$J$\,=\,3/2--1/2 and 1/2--1/2
(see Sect.\,\ref{sect:opacity}), respectively. The integrated
intensity $I$($^{13}$CN) contains the
$^{13}$CN\,$J$\,=\,1/2--1/2 and 3/2--1/2 components.
The factor of 1.082 is caused by the fact that a weak $^{13}$CN\,($F_1$, $F_2$\,=\,1--0)
hyperfine feature near 108.4\,GHz is both below our detection threshold and offset
from the considered frequency range (see \citealt{Henkel2014}).
The $^{12}$C/$^{13}$C isotope ratios derived from the CN line
ratios following Equation\,\ref{Equ:1} range from 30 to 67 with
an average of 41.6\,$\pm$\,0.2 in NGC\,253, from 24 to 62 with
an average of 38.3\,$\pm$\,0.4 in NGC\,1068, and from 6 to 44
with an average of 16.9\,$\pm$\,0.3 in NGC\,4945 (see Table\,\ref{table:ISO-ratio}).
The uncertainty of these $^{12}$C/$^{13}$C isotope ratio is $\sim$10\%,
which is mainly caused by the absolute flux calibration error of $\sim$5\%.

\section{Discussion}
\label{sect:discussion}
\subsection{Comparison to previous $^{12}$C/$^{13}$C isotopic ratio measurements}
\label{Sec:previous}
As deduced in Sect.\,\ref{sect:ISO}, with ALMA we find average $^{12}$C/$^{13}$C
isotope ratios of 41.6\,$\pm$\,0.2, 38.3\,$\pm$\,0.4, and 16.9\,$\pm$\,0.3 in the nuclear disks of
NGC\,253, NGC\,1068, and NGC\,4945, respectively
(see Table\,\ref{tab:averaged-ratio}).
We compare our ALMA measured $^{12}$C/$^{13}$C isotope ratios with previous results
obtained from single-dish observations in Table\,\ref{tab:averaged-ratio}.
For NGC\,253 and NGC\,1068, our average measured $^{12}$C/$^{13}$C isotope ratios agree well with
those obtained from the single-dish observations \citep{Henkel1993a,Henkel1993b,Henkel2014,Nakajima2018}.
However, our ALMA measured $^{12}$C/$^{13}$C isotope ratio is
significantly lower than that obtained from the single-dish
observations in NGC\,4945 \citep{Henkel1993a,Henkel1994a,Wang2004}.
This may be caused by several reasons.
Previous single-dish observations may include a lot of material from outside
the nuclear disk, i.e. from the bar extending from galactocentric radii of 100--300\,pc
and from the spiral arms even farther out (see the sketch in \citealt{Henkel2018}).
Our ALMA data cover a smaller region with galactocentric radii out to $\sim$175\,pc.
The $^{12}$CN distribution appears to be complex in NGC\,4945
(see Sect.\,\ref{sect:CN-distribution}). The typical linewidths of $^{12}$CN
are $\sim$50\,km\,s$^{-1}$ at velocity $\sim$700\,km\,s$^{-1}$ and
$\sim$150\,km\,s$^{-1}$ at velocity $\sim$450\,km\,s$^{-1}$ in
the northeastern and southwestern regions of its highly inclined
($i$\,$\sim$\,75$\degr$) nuclear disk (see our Fig.\,\ref{fig:spectra}
or Figs.\,5, 11, and 13 in \citealt{Henkel2018}).
Perhaps, the starburst is still young and the gas moving outwards through the
nuclear disk is a remnant of formerly quiescent gas highly enriched in
$^{13}$C by AGB stars through the CNO cycle (like in our Galactic center),
while $^{12}$C enrichment from young massive stars has not yet taken
over in a substantial way.

Considering the above mentioned uncertainties of our interferometric and previous single-dish
observations in the $^{12}$CN and $^{13}$CN\,$N$\,=\,1--0 transitions,
our averaged $^{12}$C/$^{13}$C isotope ratios in NGC\,253 ($\sim$400\,pc)
and NGC\,1068 ($\sim$500\,pc) confirm previous results from single-dish observations.
We conclude that the averaged $^{12}$C/$^{13}$C isotope ratios are $\sim$40--50 in
NGC\,253 and NGC\,1068 and $\sim$20--50 in NGC\,4945. The apparent discrepancy
between the single-dish and our interferometric results in NGC\,4945 will be further discussed
in Sect.\,\ref{Sec:evolution}. For NGC\,4945 ($\sim$350\,pc),
our ALMA data offer a first value for the $\sim$200\,pc sized nuclear disk, while the bar, the
inner spirals, and the nuclear $\sim$50\,pc are still waiting for a dedicated measurement.
So further observations combining our 12-m ALMA with 7-m ACA and Total Power (TP) data of
CN and eventually also other molecules and their isotopologues could provide further progress
in the determination of accurate $^{12}$C/$^{13}$C isotope ratios for the various distinct morphological components.

Presently, high resolution ($\sim$3$''$) observations of C$^{18}$O with ALMA toward NGC\,253
suggest a low $^{12}$C/$^{13}$C isotope ratio of $\sim$21$\pm$6 \citep{Martin2019},
which is a factor of $\sim$2 lower than our results obtained
from CN. We compare our ALMA measured $^{12}$CN/$^{13}$CN ratio map with their results from
$^{12}$C$^{18}$O/$^{13}$C$^{18}$O ratios in NGC\,253 (see Fig.\,3 in \citealt{Martin2019}).
It shows that the lowest $^{12}$C$^{18}$O/$^{13}$C$^{18}$O ratios are associated with the
northeastern hotspot which agrees with our results derived from $^{12}$CN/$^{13}$CN ratios,
but the distributions of $^{12}$CN/$^{13}$CN and $^{12}$C$^{18}$O/$^{13}$C$^{18}$O ratios are
slightly different in other locations. The cause(s) of these discrepancies may be different
beam sizes, different molecular distributions due to chemistry or critical densities and could
be settled by the combination of 12-m, 7-m, and TP measurements, making use of all the
instruments available at the ALMA site.

\begin{figure*}[t]
\centering
\includegraphics[width=1.00\textwidth]{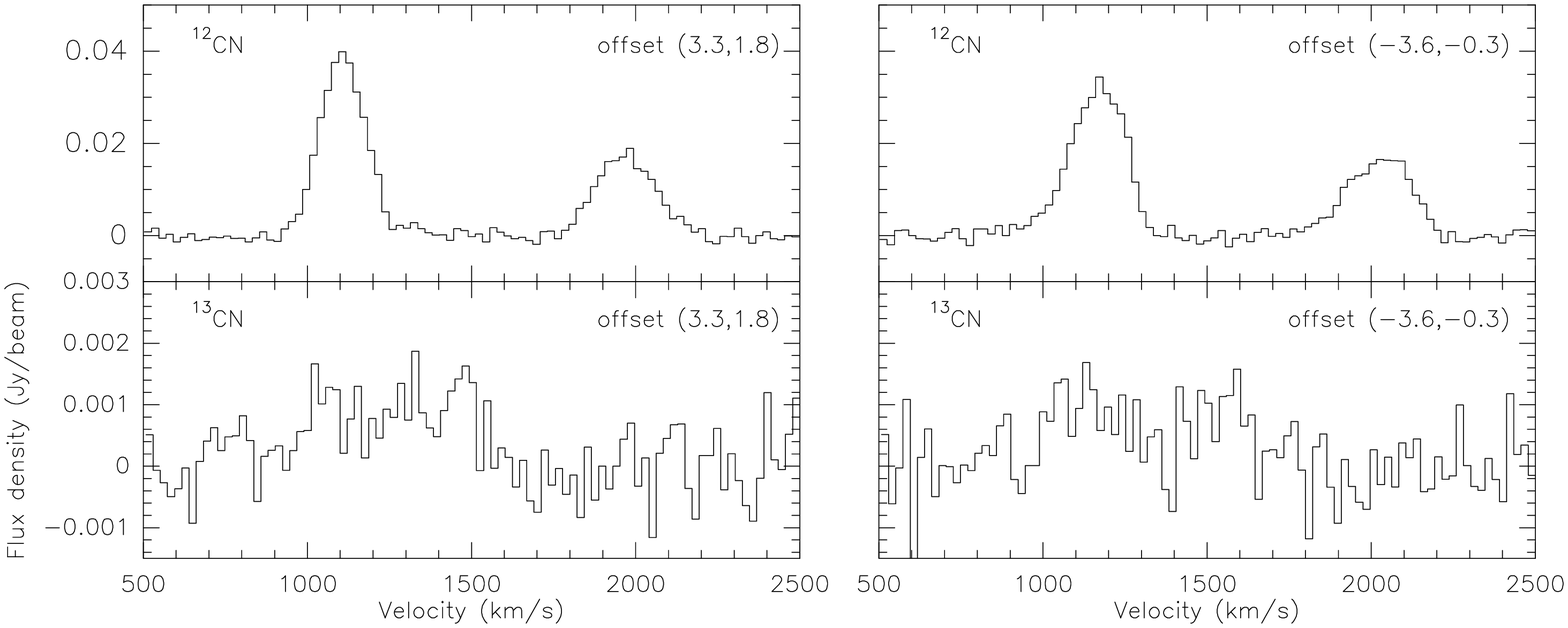}
\caption{$^{12}$CN\,$N$\,=\,1--0 (\emph{top}) and
$^{13}$CN\,$N$\,=\,1--0 (\emph{bottom}) spectra toward two outflow knots
OUT-III (\emph{left}) and OUT-II (\emph{right}) \citep{Garcia2014,Garcia2017},
respectively, in NGC\,1068 (see Sect.\,\ref{sect:Variation}).
The rest frequencies of $^{12}$CN and $^{13}$CN\,$N$\,=\,1--0 spectra are centered at
113.491 and 108.780\,GHz, respectively.}
\label{fig:Spectra-NGC1068}
\end{figure*}

\subsection{Variation of the $^{12}$C/$^{13}$C isotopic ratio}
\label{sect:Variation}
In our Galaxy, previous observations of H$_2$CO, CO, and CN indicate the presence of
a $^{12}$C/$^{13}$C isotope ratio gradient and further suggest a significant dispersion
at a given galactocentric radius (e.g., \citealt{Henkel1980,Henkel1982,Henkel1983,
Henkel1985,Henkel1994b,Langer1990,Langer1993,
Wilson1994,Savage2002,Milam2005,Giannetti2014}). This is expected in an inside-out
formation scenario for our Galaxy and in view of radial
gas streaming and potential cloud-to-cloud variations due to local supernovae
or ejecta by late-type stars (e.g., \citealt{Milam2005,Henkel2014}).
However, even in the Galactic center region isotope ratios are not everywhere
the same as recently discovered by \cite{Riquelme2010} and \cite{Zhang2015}.
The data of the former authors indicate that gas from the halo is accreted
to the disk and from the outskirts of the disk to regions closer to the
Galactic center \citep{Riquelme2010}.

In analogy to the Galactic center region, our extragalactic ALMA data also reveal
spatial variations of the $^{12}$C/$^{13}$C isotopic ratio within the
inner $\sim$400, $\sim$500, and $\sim$350\,pc sized regions
of NGC\,253, NGC\,1068, and NGC\,4945, respectively.
As mentioned in Sect.\,\ref{sect:ISO}, higher line ratios of
$^{12}$CN/$^{13}$CN indicate higher $^{12}$C/$^{13}$C isotopic ratios.
Therefore, the ratio maps can be
used as a proxy for the relative $^{12}$C/$^{13}$C isotopic ratio.
$^{12}$CN/$^{13}$CN line ratio maps and histograms of NGC\,253, NGC\,1068,
and NGC\,4945 are shown in Figs.\,\ref{fig:CN-ratios}
and \ref{fig:CN-ratios-hisgram}, respectively.
Variations of the $^{12}$CN/$^{13}$CN line ratio as mentioned in
Sect.\,\ref{sect:line-ratio} indicate that highest $^{12}$C/$^{13}$C isotopic
ratios are located in some of the outskirts of the three starburst galaxies while
lowest ratios associate with galactic central regions
and/or $^{13}$CN peak emission. The velocity components in the southwest
of NGC\,253 and NGC\,4945 \citep{Henkel2014,Henkel2018} suggest higher $^{12}$C/$^{13}$C
isotopic ratios than in the northeast. The $^{12}$C/$^{13}$C isotopic ratio
varies by factors of $\sim$2--3 in NGC\,253 and NGC\,1068 and even $\sim$7
in NGC\,4945 (see Fig.\,\ref{fig:CN-ratios-hisgram} and Table\,\ref{table:ISO-ratio}).
While these extreme differences may in part be due to low $^{13}$CN signal-to-noise ratios,
the differences are too large to be entirely a consequence of this effect
(see Table\,\ref{table:Opacity-peak}).

\begin{table}[t]
\caption{Averaged $^{12}$C/$^{13}$C ratios and comparison with previous results.}
\label{table:ISO-ratio}
\centering
\begin{tabular}
{cccccccccccc}
\hline\hline 
Source &Range &Average &Previous result &Ref. \\
\hline 
NGC\,253  & 30--67  & 41.6$\pm$0.2 &40$\pm$10    &1, 2, 3 \\
NGC\,1068 & 24--62  & 38.3$\pm$0.4 &$\sim$40$^a$ &4       \\
NGC\,4945 & 6--44   & 16.9$\pm$0.3 &40--50       &1, 5, 6 \\
\hline 
\end{tabular}
\tablebib{(1) \cite{Henkel1993a}; (2) \cite{Henkel1993b}; (3) \cite{Henkel2014}; (4) \cite{Nakajima2018};
(5) \cite{Henkel1994a}; (6) \cite{Wang2004}.}
\tablefoot{$^a$ The $^{12}$C/$^{13}$C ratio is derived from the uncorrected integrated intensity
ratio $I$($^{12}$CN)/$I$($^{13}$CN) taken from \cite{Nakajima2018} (see Sect.\,\ref{sect:Spatial-filtering}).
Calibration uncertainties are not considered here but are discussed in Sect.\,\ref{sect:ISO}.}
\label{tab:averaged-ratio}
\end{table}

For NGC\,253, integrated intensities of $^{13}$CN are highest in two hotspots,
located symmetrically with respect to the nucleus, one in the northeast and
the other one in the southwest. Lowest $^{12}$C/$^{13}$C isotopic ratios associate
with these two hotspots (see Fig.\,\ref{fig:CN-ratios}) corresponding
to centimeter and millimeter continuum peak emission
(e.g., \citealt{Turner1985,Ulvestad1997,Sakamoto2011,Krips2016,Mangum2019}).
The gas at these locations may be CNO-processed by intermediate
mass stars in the more distant past. Two locations with low $^{12}$C/$^{13}$C isotopic
ratios in the northeast (3.3\arcsec,\,1.8\arcsec) (offsets relative to our reference position;
see Table\,\ref{table:Observational_parameters}) and southwest (--3.6\arcsec,\,--0.3\arcsec)
of NGC\,1068 (see Figs.\,\ref{fig:CN-ratios} and \ref{fig:Spectra-NGC1068}, and Table\,\ref{table:Opacity-peak})
associate with two outflow knots OUT-III and II \citep{Garcia2014,Garcia2017}, respectively,
so that the highly processed gas has moved away from the center, while less processed gas may be
infalling from the outer part of NGC\,1068 into (at least the outer part of) the circum-nuclear
disk (CND) through both the halo and the bar, and the CND is dominated by the outflowing motion.
The decline of the $^{12}$C/$^{13}$C isotopic ratio in these locations
may be influenced by the outflow from the AGN and/or by processed material near the AGN.
One should note that the two outflow knots OUT-II and III are identified with $\sim$1\arcsec resolution
C$_2$H data by \cite{Garcia2017}, which corresponds to a $\sim$4.5 times
higher resolution (with respect to beam area) than the CN data presented here. Beam dilution
effects may not be negligible.
Low $^{12}$C/$^{13}$C isotopic ratios in the northeast of NGC\,4945 show similar
values as those in the nuclear region (see Fig.\,\ref{fig:CN-ratios}),
which indicates that it may be strongly affected by highly processed outflowing material from
the nuclear region (for this, see \citealt{Henkel2018}) having undergone substantial star formation in the past.

\subsection{The $^{12}$C/$^{13}$C isotopic ratio evolution in starburst galaxies}
\label{Sec:evolution}
The $^{12}$C/$^{13}$C isotope ratio is a useful probe of the chemical
evolution of galaxies (e.g., \citealt{Milam2005,Martin2010,Henkel2014}).
It is believed to be a direct measure of primary to secondary nuclear processing \citep{Wilson1994}.
The $^{12}$C/$^{13}$C isotope ratio is expected to decline with time
(e.g., \citealt{Henkel1993a,Prantzos1996,Hughes2008,Martin2010,Henkel2014,Romano2017}).
This leads to the very low ratios in our Galactic center region.
However, in case of a starburst, triggered by a bar or by a merger,
gas from outside with higher ratios is flowing into a galaxy's central region,
enhancing the $^{12}$C/$^{13}$C isotope ratio. A few Myr after the start
of a starburst, this effect will be strengthened by the ejecta from massive stars.
A top-heavy stellar initial mass function could make this effect even more pronounced
(e.g., \citealt{Henkel1993a,Romano2017,Zhang2018}).

The inflow scenario discussed for starbursts may even lead to higher $^{12}$C/$^{13}$C ratios
for ultraluminous infrared galaxies (ULIRGs) since such objects have more powerful inflows
(e.g., \citealt{Toyouchi2015,Yabe2015,Falstad2017}). Indeed, ULIRGs have not only a higher $^{12}$C/$^{13}$C ratio,
but are also deviating from the canonical mass-metallicity relation in the sense of having
a lower metallicity for their mass or a higher mass for their metallicity \citep{Pereira2017}.
More moderate starburst galaxies have experienced less inflow,
which may suggest lower $^{12}$C/$^{13}$C isotope ratios.
In comparison, our Galaxy shows only weak signs of inflow (e.g., \citealt{Morris1996,Riquelme2010}),
which is reflected in the lower $^{12}$C/$^{13}$C isotope ratios measured in its central molecular zone.
The CN data presented here indicate that the averaged $^{12}$C/$^{13}$C isotope ratios in
the nuclear regions of NGC\,253 and NGC\,1068 are higher than in our Galactic center region.
Nevertheless, the averaged $^{12}$C/$^{13}$C isotope ratios in our selected
nearby starburst galaxies are lower than previous observational results
in the well-studied ULIRGs Arp\,220 and Mrk\,231 ($\sim$100; \citealt{Gonzalez2012,Henkel2014}),
and also in the high-$z$ Cloverleaf ULIRG/QSO ($\gtrsim$100; \citealt{Henkel2010}).
This confirms the trend of declining $^{12}$C/$^{13}$C values with time
and metallicity proposed by \cite{Henkel2014}.

Measurements of the $^{12}$C/$^{13}$C isotope ratio based on CN lines have the potential to
reveal the degree of gas processing in the nuclear regions of starburst galaxies.
More galaxies are needed to study nucleosynthesis, to constrain galaxy dynamics, and to
discriminate between different evolutionary stages to follow in more detail the secular
decline of $^{12}$C/$^{13}$C ratios, occasionally interrupted by inflow and starburst
activity. With ALMA it is possible to extend such studies to objects at greater distances.

\section{Summary}
\label{sect:summary}
We have measured the $^{12}$C/$^{13}$C isotopic ratio in the nuclear regions of three nearby
starburst galaxies NGC\,253, NGC\,1068, and NGC\,4945  making use of the $^{12}$CN
and $^{13}$CN\,$N$\,=\,1--0 lines in the ALMA Band 3 at frequencies near 110\,GHz.
The main results are the following:
\begin{enumerate}
\item
The $^{12}$C/$^{13}$C isotopic ratios derived from the $^{12}$CN and $^{13}$CN
line ratios range from 30 to 67 with an average of 41.6\,$\pm$\,0.2
in NGC\,253, from 24 to 62 with an average of 38.3\,$\pm$\,0.4 in NGC\,1068,
and from 6 to 44 with an average of 16.9\,$\pm$\,0.3 in NGC\,4945.
The $^{12}$C/$^{13}$C isotopic ratios vary by factors of $\sim$2--3
in NGC\,253 ($\sim$400\,pc) and NGC\,1068 ($\sim$500\,pc) and $\sim$7 in NGC\,4945 ($\sim$350\,pc).
The large scatter of values, particularly in NGC 4945, is certainly in part a consequence
of the limited sensitivity of our data. Neverthelss, the variations are too large to be
only caused by this effect, suggesting the presence of real variations as they have
recently been found in our Galactic center region.

\item
The highest $^{12}$C/$^{13}$C isotopic ratios are located in the outskirts of the three
starburst galaxies' nuclear regions. The lowest ratios are associated with the northeastern and southwestern
molecular peaks of NGC\,253, the northeastern and southwestern edge of the mapped region
in NGC\,1068, and the very center of NGC\,4945.

\item
The measured $^{12}$C/$^{13}$C isotopic ratios in NGC\,1068 indicate that
the highly processed gas has moved away from the center and less processed gas
may be infalling from the outer part of NGC\,1068 into the CND through
both the halo and the bar.

\item
Low $^{12}$C/$^{13}$C isotopic ratios in the central regions of these starburst
galaxies indicate the presence of highly processed material.

\item
Our results agree with the scenario of $^{12}$C/$^{13}$C ratios slowly decreasing
with time in galaxies.

\end{enumerate}

\begin{acknowledgements}
The authors thank the anonymous referee for helpful comments.
This work acknowledges support by The Heaven Lake Hundred-Talent Program
of Xinjiang Uygur Autonomous Region of China, The National Natural Science
Foundation of China under grant 11433008, and The Collaborative Research
Council 956, subproject A6, funded by the Deutsche Forschungsgemeinschaft (DFG).
C.\,H. acknowledges support by Chinese Academy of Sciences President's International
Fellowship Initiative under Grant No.\,2019VMA0039.
This paper makes use of the following ALMA data:
ADS/JAO.ALMA\#2011.0.00172.S and 2013.1.01151.S.
ALMA is a partnership of ESO (representing its member states),
NSF (USA) and NINS (Japan), together with NRC (Canada), NSC and
ASIAA (Taiwan), and KASI (Republic of Korea), in cooperation
with the Republic of Chile. The Joint ALMA Observatory is
operated by ESO, AUI/NRAO and NAOJ.
This research has made use of the NASA/IPAC Extragalactic Database (NED)
which is operated by the Jet Propulsion Laboratory, California Institute of
Technology, under contract with the National Aeronautics and Space Administration.
This research has used NASA's Astrophysical Data System (ADS).
\end{acknowledgements}

\end{document}